\documentclass{llncs}
\usepackage{graphicx}
\usepackage{caption}
\usepackage{subcaption}
\usepackage{mathtools}
\usepackage{multirow}
\usepackage[utf8x]{inputenc} 
\usepackage{xcolor}
\usepackage{pgfplots}
\usepackage{url}

\begin{document}
	
	\title{A Scalable Document-based Architecture \\for Text Analysis}
	
	\author{Ciprian-Octavian Truică$^1$, J{\'e}r{\^o}me Darmont$^2$, Julien Velcin$^2$
	}
	
	\institute{
		$^{1}$ Computer Science and Engineering Department, Faculty of Automatic Control \\and Computers, University Politehnica of Bucharest, Bucharest, Romania\\
		\texttt{ciprian.truica@cs.pub.ro}\\
		$^{2}$ Univ Lyon, Lumi{\`e}re, ERIC EA 3083, Lyon, France\\
		\texttt{\{jerome.darmont,julien.velcin\}@univ-lyon2.fr}
	}
	\maketitle
	
	\begin{abstract}
		Analyzing textual data is a very challenging task because of the huge volume of data generated daily. Fundamental issues in text analysis include the lack of structure in document datasets, the need for various preprocessing steps 
and performance and scaling issues. Existing text analysis architectures partly solve these issues, providing restrictive data schemas, addressing only one aspect of text preprocessing and focusing on one single task when dealing with performance optimization. 
Thus, we propose in this paper a new generic text analysis architecture, where document structure is flexible, many preprocessing techniques are integrated and textual datasets are indexed for efficient access. We implement our conceptual architecture using both a relational and a document-oriented database. Our experiments demonstrate the feasibility of our approach and the superiority of the document-oriented logical and physical implementation.
	\end{abstract}
	
	
	\section{Introduction}
	
	A vast amount of textual data is generated daily and it is really challenging to develop efficient models and systems to enhance processing performance while doing accurate text analysis. The most fundamental challenges when working with large volumes of heterogeneous text datasets include the lack of structure of textual corpora, the various required preprocessing steps, 
the need for efficient access and the ability to scale up. 
	
	Structural issues may be addressed by resorting to textual data warehousing and On-Line Analytical Processing (OLAP). However, such approaches only partially solve the problem because they use a structured schema 
that falls short when applied to large, heterogeneous volumes of data. Moreover, using a predefined schema makes 
them extremely dataset-specific. 

	Moreover, when dealing with textual data, we distinguish different preprocessing levels: quite basic operations (e.g., cleaning HTML tags, tokenization, language identification); intermediate operations (e.g., stemming, lemmatization, indexing); and advanced operations (e.g., part of speech tagging, named entity recognition, topic modeling). Each complexity layer in this process requires the previous layer and all operations must remain tractable in terms of memory and CPU time. To the best of our knowledge, no text analysis tool implements all layers, nor any processing workflow.

	Finally, when working on performance and scaling issues, state-of-the-art research focuses on one aspect of text analysis, e.g., aggregation, top-k keyword extraction 
and text indexing. 
However, text processing techniques used in a single application may be many and, as we mention above, interdependent.
	
	Hence, we present in this paper a scalable text analysis architecture that addresses all these issues. More precisely, we deal with the lack of structure by adopting a novel generic, document-oriented data model that allows storing heterogeneous textural corpora with no predefined structure. We also integrate in our framework all the preprocessing methods that are useful for information retrieval, data mining, text analysis and knowledge discovery. We also propose a new compact data structure to minimize index storage space and the response time of 
create, read, update and delete (CRUD) operations. Such indexes benefit to text preprocessing, querying and further analysis, and adequately contribute to global scaling.
	
	The remainder of this paper is organized as follows. In Section~\ref{sec:RelatedWorks}, we discuss related works. 
In Section~\ref{sec:ProposedApproachAndImplementation}, we present the architecture and implementation of our approach. 
In Section~\ref{sec:ExperimentalValidation}, we experimentally validate our proposal. 
In Section~\ref{sec:Conclusion}, we finally conclude this paper and hint at future research.
	
	\section{Related Works}\label{sec:RelatedWorks}
	\subsection{Text Cubes and OLAP}
	Extensive work on information retrieval (IR) and text analysis have been done using OLAP. 
	Most proposals use Text Cubes for 
OLAPing multidimensional text databases~\cite{DZhang2009}. Lin et al. focus on optimizing query processing and reducing storage costs of Text Cubes~\cite{CXLin2008}. They experimentally 
show that average query time and storage cost are related to a cube's number of dimensions. Zhang et al. use Text Cubes for topic modeling~\cite{DZhang2009} and 
experimentally show that their approach is much faster than 
computing each topic cube from scratch. Finally, Ding et al. address the problem of keyword search and top-k document ranking using Text Cubes~\cite{BDing2011}. Their algorithms perform well in terms of query response and memory cost when the number of search terms is small. 
	
	Ben Kraiem et al. propose a generic multidimensional model for OLAP on tweets~\cite{MBenKraiem2014}. Their experiments show some promising results for knowledge discovery when applying OLAP on a small 
corpus, but 
query performance decreases when data volume increases. Bringay et al. propose a data warehouse model to analyze large volumes of tweets~\cite{SBringay2011}. They introduce different operators to identify trends using the top-k most significant words over a period of time for a specific geographical location, as well as the impact of hierarchies on such operators. 
Unfortunately, no time performance and storage cost analysis is provided.

	In conclusion, research done so far on text analysis and OLAP focuses on small, structured datasets and 
	scaling up is not guaranteed.
	
	\subsection{Text Preprocessing and Analysis}
	Managing morphological variation of search terms in IR has been quite extensively studied~\cite{KKettunen2005,AGJivani2011}. The main successful methods are stemming~\cite{DSharma2012} and lemmatization, which are used to optimize search, minimize the space allocated to inverted indexes (Section~\ref{sec:docindex}) and, in the case of lemmatization, to add linguistic information. Lemmatization is useful for different types of advanced text analysis, e.g., named entity recognition, automatic domain specific multi-term extraction and part of speech (PoS) tagging. Moreover, lemmatization is easier of use than stemming, saves storage and improves retrieval performance~\cite{KKettunen2005}.
	
	Topic modeling is a statistical model for discovering hidden themes that occur in a collections of documents. In recent years, it has been extensively studied, showing the usefulness of analyzing latent topics and discovering topic patterns~\cite{DMBlei2003,
YWTeh2006}. 
	Popular approaches for topic modeling are latent semantic indexing (LSI)~\cite{SCDeerwester1990}, 
	latent Dirichlet allocation (LDA)~\cite{DMBlei2003}, the non-parametric extension hierarchical Dirichlet process (HDP)~\cite{YWTeh2006} and non-negative matrix factorization (NMF)~\cite{SArora2013}.

	\subsection{Document Indexing}\label{sec:docindex}
	Inverted indexes are data structures used in search engines, whose main purpose is to optimize query response speed. Basic inverted indexes store terms, a list of documents where each term appears and a weight. Weight measures the number of occurrences of the term in a document, e.g., raw term frequency/word co-occurrence (TF), normalized Term Frequency (TF$_n$), etc. In the various methods for managing inverted indexes, great emphasis is put on storage space reduction. For instance, a pruning algorithm based on term frequency-inverse document frequency (TF*IDF) can be used to minimize index size~\cite{SKVishwakarma2014}. Yet, updating an inverted index is also a problem, because it is dependent on  documents. The index must indeed be updated each time documents are added or deleted. 
	
	\section{Proposed Approach and Implementation}\label{sec:ProposedApproachAndImplementation}
	
	\subsection{Approach Overview}
\begin{figure}
		\centering
		\includegraphics[width=10cm]{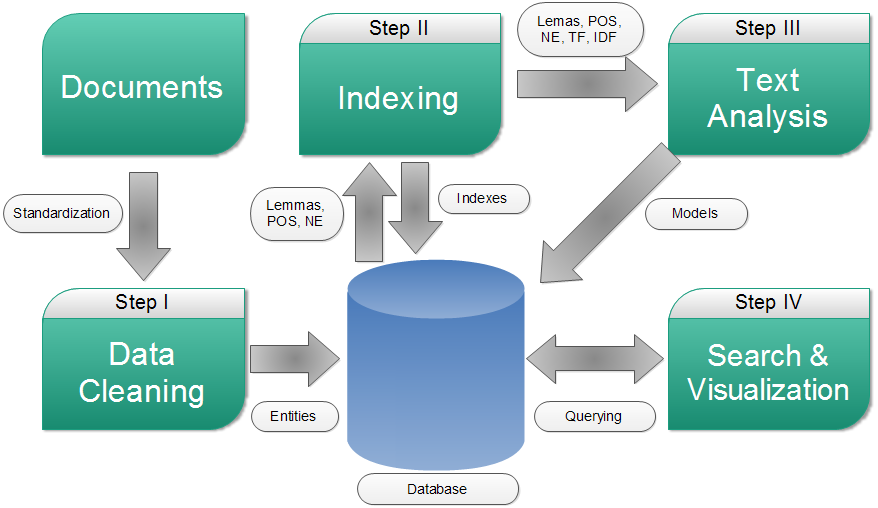}
		\caption{System architecture}
		\label{fig:architecture}
	\end{figure}

	The approach we propose (Figure~\ref{fig:architecture}) is subdivided into four steps: 1) clean and preprocess documents using natural language processing (NLP) and store the information in a database; 2) construct indexes; 3) analyze data, e.g., with topic modeling, etc.; 4) query and search data, extract top-k most relevant documents, create visualizations and analyses.
	We construct the inverted index, vocabulary, PoS and named entities (NE) indexes during the index construction step. Indexes may be used afterward by data mining, text analysis, search and visualization. The search engine sorts documents based on a ranking function (e.g., TF*IDF, Okapi or BM25) to extract the top-k documents. 

	To implement our document-oriented approach, we quite naturally rely on a document-oriented database management system (DODBMS). DODBMSs are a class of NoSQL systems that aim to store, manage and process data using a semi-structured model. DODBMSs encapsulate data in collections of documents~\cite{JHan2011}. 
A document can contain other nested documents, which turns out to be very flexible~\cite{COTruica2015a}.

		One feature of DODBMSs is that they are often optimized for create and read operations, while offering reduced functionality for update and delete queries. DODBMSs are designed to work with large amounts of data and the main focus is on the efficiency of data storage, access and analysis~\cite{ERedmond2012}.
	Another key feature of DODBMSs is the distribution of data across multiple sites. 
In particular, DODBMSs can horizontally scale CRUD operations throughput~\cite{RCattell2011}. 
	Moreover, decentralized data stores provide good mechanisms for fail-over, removing the single point of failure, due to their scalability and flexibility~\cite{RHecht2011}.
	
	We selected MongoDB as our DODBMS, since it beats the best mean time performances for CRUD operations both in single and distributed environments \cite{COTruica2015a}. Moreover, we also implemented our approach with PostgreSQL, to provide a point of comparison with a well-established, efficient relational database management system (RDBMS) 
	(Section~\ref{sec:ExperimentalValidation}).
	
	\subsection{Data Models}
\begin{figure}
		\centering
		\includegraphics[width=12cm]{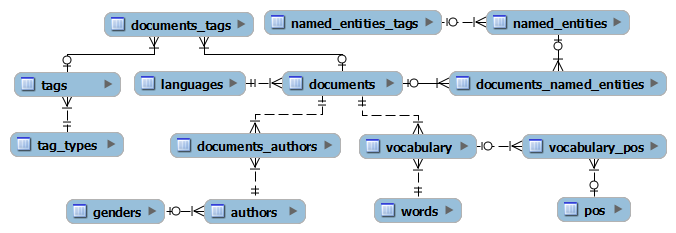}
		\caption{Conceptual model}
		\label{fig:postgredb}
	\end{figure}

	We design a generic model to store heterogeneous text data using a data warehouse snowflake schema 
    (Figure~\ref{fig:postgredb}). The central component of the model is the \emph{documents} entity, where we store basic information and metadata about a document, e.g., timestamp, title, raw, clean and lemmatized text, etc. The \emph{document\_tags} entity is used to store metadata represented by tags, which can be existing tags, hashtags or at tags. 
	The \emph{vocabulary} entity links documents to information extracted or inferred from the text, which helps enhancing metadata with different weights and tags, e.g., PoS, TF, TF$_n$, lemmas, etc. The \emph{named\_entities} entity stores all the information about entities automatically extracted from the original corpus.
	
	The DODBMS schemaless design  takes all the information presented in the relational schema and stores it for each document in a record of the collection. 
Using this design, all one-to-many and many-to-many relationships become either vectors (e.g., \emph{hashtags}, \emph{at tags}) or nested documents (e.g., \emph{words}, \emph{named\_entities}). Where the information is not present, these vectors and nested documents may be missing thanks to the flexibility of schemaless database design. A problem that arises is duplication, as multiple records can bear the same metadata, since all the information for a document is stored in one single record. The \emph{vocabulary} entity is constructed as a separate collection. This entity is constructed dynamically, taking  user input constraints into account, e.g. date, tags, search words, named-entities.
	
	
	Interaction with the database is achieved through CRUD operations, aggregation functions and views. We use read operations for information extraction and data visualization. Aggregation functions are used for constructing indexes, searching and preprocessing data for text analysis. We make use of MapReduce for this purpose when using the document-oriented database architecture. Dynamically materialized cubes are constructed using views with aggregation functions, fine-graining query results using different measures, e.g., timestamps, locations, lemmas, tags, named entities.
	
	\subsection{Text Preprocessing}
	
	The data cleaning module serves three functions: 1) corpus standardization, 2) text preprocessing using NLP to enrich data, and 3) entity creation and information insertion into the database. 
	
	The entire corpus is standardized by determining all the fields of a document, including metadata and the labels of \emph{documents}. 
    Then, during the preprocessing step, the following techniques are applied: 1) text cleaning by removing HTML/XML tags and scripts; 2) language identification; 3) expanding contractions; 4) extracting features, e.g., PoS, lemmas and named entities; 5) removing stop words and punctuation; 6) computing term weights. We use a multithreading architecture for data cleaning to cope with large data volumes   and scale up vertically. At the end of each thread, the information is stored in a dictionary, together with other metadata. We choose to use asynchronous threads because, after a worker thread finishes, a new job can be assigned to it without waiting for the other worker threads to finish. This is made possible because each task is independent. At the end of this step, a record of the \emph{documents} collection is created and inserted into the database. The record contains all labels from the first step and the information extracted using NLP from the second one. 
	
	In the DODBMS implementation, 
	a record stores all the information because its attributes are created dynamically. In contrast, the RDBMS architecture can only store predefined fields due to its rigid schema. Thus, undefined fields  are omitted. 
	
		The RDBMS approach merges the data cleaning step with the index construction step, because many-to-many relationships between entities, translated as bridge tables, are indexes as well. We could not use a multithreading approach here because information could be lost. Multiple threads could indeed check at the same time whether the information is present and receive a negative response. A constraint violation error could appear and the transaction terminate by a \emph{rollback}. If constraints are missing, then duplicate information could appear and this would impact text analysis.
	
	\subsection{Index Management}
    We propose several indexes for document aggregation, search, extraction of the top-k most signification terms and text analysis, e.g., topic modeling, document clustering. These new indexing structures minimize storage costs and maximize the time performance of CRUD operations.
    
\begin{figure}
		\centering
		\includegraphics[width=11cm]{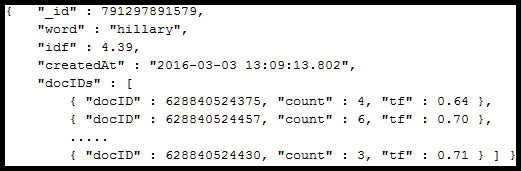}
		\caption{Vocabulary index structure}
		\label{fig:voc_ex}
	\end{figure}

	Index construction in the DODBMS architecture is done using the MapReduce framework. Four indexes are created: 1) an \emph{inverted index} that stores, for each term, a list of corresponding documents; 2) a \emph{vocabulary}, a novel inverted index with additional information for each term in the corpus, e.g., list of documents where the term is found, the TF and TF$_n$ of the term for a document and IDF; 3) a PoS index that stores the part of speech of each term; 4) a named-entity index used for storing named entities. There are no integrity constraints between these collections to improve query response time. Moreover, the structure proposed for the \emph{vocabulary} facilitates query response time, aggregation and search (Figure~\ref{fig:voc_ex}).
			MapReduce is used to construct all indexes. It is also central in aggregation queries needed by the search algorithms. To improve  index construction and query response times, we horizontally scale the database, and by doing so add more MapReduce worker. 
			
			In the RDBMS architecture, indexes are the bridge tables translating many-to-many relationships between entities. The \emph{vocabulary} is the bridge table between the \emph{documents} table and the \emph{words} table. The PoS index is the bridge table between the \emph{vocabulary} table and the \emph{pos} table. In this case, the index also contains the TF and IDF of each term.
			
			The number of entries in the indexes constructed for the DODBMS is equal to the number of terms in the entire corpus. In the RDBMS, the inverted index has more entries, i.e., $\sum\nolimits_{d \in D} \mid t : t \in d \mid$ , where \emph{D} is the  corpus and $\mid t : t \in d\mid$ is the number of distinct terms that appear in document $d$.
			
	Updating indexes in the DODBMS is based on document insertion date. The update method  we use constructs an intermediary index for new documents, and then it updates the primary index by appending the new documents' ID and TF to existing labels. Then, the IDF of each term is updated for the whole index. When documents are deleted, we apply a bulk delete operation. In this case, a list of deleted document IDs is stored, which helps update the index structure by removing the deleted documents and then updating the IDF of each term.
	
	Updating indexes in the RDBMS implementation is easier thanks to the database's structure. When documents are added, indexes are automatically updated based on the insertion date of the last added documents. When documents are removed, the corresponding index entries are also removed. For both operations, the IDF of each term must be recalculated.

	\section{Experimental Validation}\label{sec:ExperimentalValidation}
	
	In this section, we test each step of our approach and we compare the results achieved by the two instances we developed, i.e., the DODBMS version implemented with MongoDB and the RDBMS version implemented with PostgreSQL. Tests are done using a news corpus consisting of 110,000 articles\footnote{\url{http://www.corpora.heliohost.org}}, a corpus of 5,000,000 tweets\footnote{Collected with Twitter's tools at \url{https://dev.twitter.com}} and a scientific corpus of 20,000 abstracts from \cite{JTang2012}. 
	The size of these corpora, some would argue, is rather small with respect to Big Data. Yet, it is sufficient to illustrate our architecture's good time performance. Text analysis is indeed not usually done on large corpora. Moreover, other corpora used in the literature are smaller, e.g., 3,000 documents~\cite{DZhang2009}, 2,013 records~\cite{CXLin2008}, 
	65,333 tweets~\cite{MBenKraiem2014}, 1,801,810 tweets~\cite{SBringay2011}. 
    	
		Our architecture can be deployed in a cloud environment if all the requirements are met, i.e., if Python packages, PostgreSQL, and MongoDB are available. Tests are done on machines that reside in an OpenStack private cloud platform. We purposely selected this hardware architecture and dataset sizes to show that our architecture can achieve good performance even  on end-user workstations, as  it is sometimes not desirable to send data online due to privacy issues. Moreover,  end-users presumably cannot afford very powerful, parallel computers.
	
	\subsection{News Articles Corpus Experiments}
	The first set of experiments are done using two computers with the same hardware configuration: 4 GB RAM and 1 CPU with two 2.2 GHz cores. We choose this hardware architecture to show that our method gives good results on simple  computers.
	Using the initial news articles corpus, seven corpora are created consisting of 100 
	to 110,000 documents. 
	They are referred to as Corpus~$i, i \in \{1, 2, ... , 7\}$. For comparison reasons, 
	experiments are done using a single-thread approach.
	
	Figure~\ref{fig:populatedb} presents the average time (in seconds) for populating the databases. 
    Duplicate documents are removed in this step. This is done by checking whether an article already exists in the database based on its title. If the document does not exist, then a new record is added. Otherwise,  tags are verified so that metadata are not omitted, as the same article could have more tags for different instances found in the corpus. The second set of tests evaluates the efficiency of text cleaning and index construction (Figure~\ref{fig:text_preprocessing}).
	
	\pgfplotsset{height=0.70\columnwidth, width=1.00\columnwidth}
	\begin{figure}[hbt]
		\centering
		\begin{subfigure}{.5\textwidth}
			\centering
			\begin{tikzpicture}[]
			\begin{axis}[
			bar width=3pt,
			legend style={anchor=north west, legend cell align=left, at={(0.0,1.0)}},
			ybar = 0pt,
			xmin = 0,
			xmax = 8,
			xtick = {1,2,3,4,5,6,7},
			xtick align=inside,
			xlabel={Corpus},
			ylabel={Time in seconds}
			]
			\addplot+[color=blue] table [x=corpus, y=mongodb]  {populate_db.txt};
			\addlegendentry{MongoDB}
			\addplot+[color=red] table [x=corpus, y=postgresql]  {populate_db.txt};
			\addlegendentry{PostgreSQL}
			\end{axis}
			\end{tikzpicture}
			\caption{Data insertion comparison}
			\label{fig:populatedb}
		\end{subfigure}%
		\begin{subfigure}{.5\textwidth}
			\centering
			\begin{tikzpicture}[]
			\begin{axis}[
			bar width=3pt,
			legend style={anchor=north west, legend cell align=left, at={(0.0,1.0)}},
			ybar = 0pt,
			xmin = 0,
			xmax = 8,
			xtick = {1,2,3,4,5,6,7},
			xtick align=inside,
			xlabel={Corpus},
			ytick \empty, 
			extra y ticks = {0,25000,50000,75000, 100000},
			extra y tick labels = {0,5k,10k,15k,20k,25k,30k,35k},
			ylabel={Time in seconds}
			]
			\addplot+[color=blue] table [x=corpus, y=mongodb]  {text_cleaning.txt};
			\addlegendentry{MongoDB}
			\addplot+[color=red] table [x=corpus, y=postgresql]  {text_cleaning.txt};
			\addlegendentry{PostgreSQL}
			\end{axis}
			\end{tikzpicture}
			\caption{Text preprocessing comparison}
			\label{fig:text_preprocessing}
		\end{subfigure}
		\begin{subfigure}{.5\textwidth}
			\centering
			\begin{tikzpicture}[]
			\begin{axis}[
			bar width=3pt,
			legend style={anchor=north west, legend cell align=left, at={(0.0,1.0)}},
			ybar = 0pt,
			xmin = 0,
			xmax = 8,
			xtick = {1,2,3,4,5,6,7},
			xtick align=inside,
			xlabel={Corpus},
			ytick \empty, 
			extra y ticks = {0,500,1000,1500},
			extra y tick labels = {0,500,1\,000,1\,500},
			ylabel={Size in MB}
			]
			\addplot+[color=blue] table [x=corpus, y=mongodb]  {storage.txt};
			\addlegendentry{MongoDB}
			\addplot+[color=red] table [x=corpus, y=postgresql]  {storage.txt};
			\addlegendentry{PostgreSQL}
			\end{axis}
			\end{tikzpicture}
			\caption{Storage comparison}
			\label{fig:storage}
		\end{subfigure}%
		\begin{subfigure}{.5\textwidth}
			\centering
			\begin{tikzpicture}[]
			\begin{axis}[
			bar width=3pt,
			legend style={anchor=north west, legend cell align=left, at={(0.0,1.0)}},
			ybar = 0pt,
			xmin = 0,
			xmax = 6,
			ymin = 0,
			ymax = 5,
			xtick = {1,2,3,4,5},
			xtick align=inside,
			xlabel={No. terms},
			ylabel={Time in seconds},
			extra y ticks = {0,1,2,3,4},
			extra y tick labels = {0,1,2,3,4}
			]
			\addplot+[color=blue] table [x=terms, y=mongodb]  {search.txt};
			\addlegendentry{MongoDB}
			\addplot+[color=red] table [x=terms, y=postgresql]  {search.txt};
			\addlegendentry{PostgreSQL}
			\end{axis}
			\end{tikzpicture}
			\caption{Search comparison}
			\label{fig:search_performance}
		\end{subfigure}
		\caption{Performance comparison}
	\end{figure}
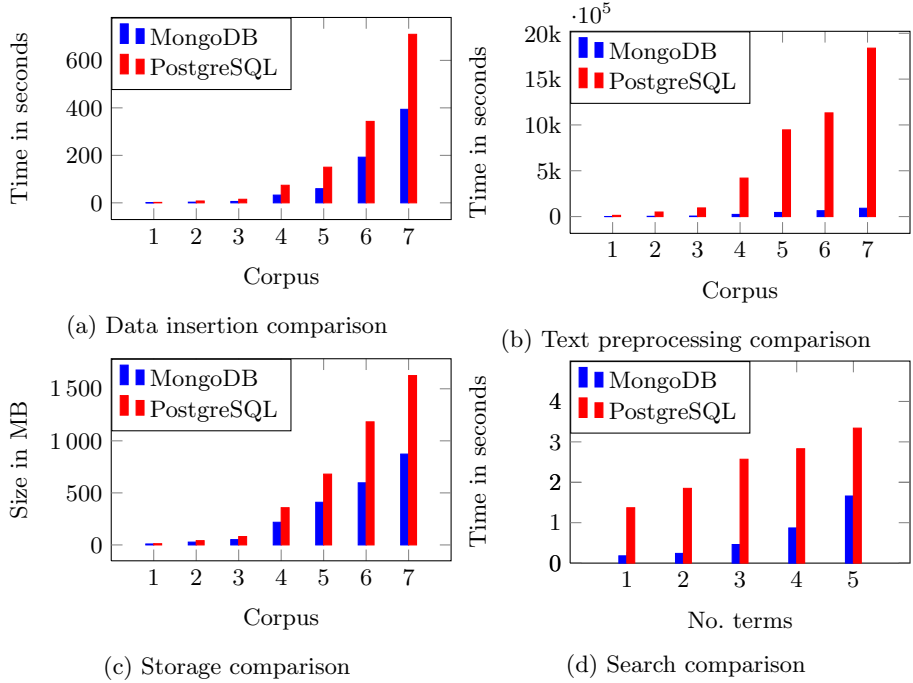
	
	Figure~\ref{fig:storage} shows the total storage space (in MB) for all corpora. To respect database normalization in PostgreSQL, bridge tables materializing many-to-many relations have to be added. In MongoDB, such relationships translate into vectors or nested documents inside collections. For example, the \emph{documents} collection contains the authors table as an array of nested documents and the tags table as an array. This brings the issue of duplicates, as we may have the same tags for different document that would be stored in each element of the collection. 
	However, it is a small cost to pay as, using this structure, joins are removed, whereas join is the costliest operation in RDBMSs.
	
Experimental results show that MongoDB efficiently stores the data, minimizing storage space by 30\% with respect to PostgreSQL. Moreover, based on the number of records in each collection, from a computational point of view, a select operation performed on a smaller entity shows faster response times than one performed on an entity with a lot of records. For example, it is faster to query the \emph{vocabulary} collection than to interrogate the \emph{vocabulary} table, because the table contains more records than the collection.
	
	\begin{table}[hbt]
		\centering
		\caption{Index building and updating in MongoDB (seconds)}
		\begin{subtable}{1\linewidth}
			\centering
			\caption{Text cleaning and index construction}
			\label{tab:create_idx}
			\begin{tabular}{|l|r|r|r|r|}
				\hline
				\multicolumn{1}{|c|}{\textbf{Corpora}} & \multicolumn{1}{c|}{\textbf{Clean Text}} & \multicolumn{1}{c|}{\textbf{PoS Index}} & \multicolumn{1}{c|}{\textbf{Inverted Index}} & \multicolumn{1}{c|}{\textbf{Vocabulary}} \\ \hline
				Corpus 1 & 25.36 & 4.95 & 5.94 & 6.57 \\ \hline
				Corpus 2 & 103.74 & 12.63 & 16.15 & 19.26 \\ \hline
				Corpus 3 & 204.91 & 20.10 & 25.36 & 32.95 \\ \hline
				Corpus 4 & 934.56 & 57.74 & 81.46 & 117.15 \\ \hline
				Corpus 5 & 1\,805.27 & 103.38 & 141.81 & 209.14 \\ \hline
				Corpus 6 & 2\,524.33 & 148.69 & 212.16 & 312.61 \\ \hline
				Corpus 7 & 3\,564.07 & 216.34 & 311.88 & 461.88 \\ \hline
			\end{tabular}
		\end{subtable}
		\begin{subtable}{1\linewidth}
			\centering
			\caption{After new documents are added}
			\label{tab:update_idx_add}
			\begin{tabular}{|c|c|l|c|l|c|l|}
				\hline
				\multirow{2}{*}{\textbf{\begin{tabular}[c]{@{}c@{}}No.\\ Documents\end{tabular}}} & \multicolumn{2}{c|}{\textbf{Inverted Index}} & \multicolumn{2}{c|}{\textbf{Vocabulary}} & \multicolumn{2}{c|}{\textbf{PoS Index}} \\ \cline{2-7} 
				& \textbf{Update} & \textbf{Rebuild} & \textbf{Update} & \textbf{Rebuild} & \textbf{Update} & \textbf{Rebuild} \\ \hline
				\multicolumn{1}{|l|}{500} & {\bf 37.65} & 123.63 & 1\,163.95 & {\bf 189.62} & {\bf 33.53} & 91.52 \\ \hline
				\multicolumn{1}{|l|}{1\,000} & {\bf 76.54} & 126.16 & 1\,214.20 & {\bf 208.47} & {\bf 67.79} & 93.08 \\ \hline
				\multicolumn{1}{|l|}{2\,500} & 144.56 & {\bf 126.07} & 1\,303.89 & {\bf 204.18} & 129.44 & {\bf 95.24} \\ \hline
				\multicolumn{1}{|l|}{5\,000} & 201.26 & {\bf 130.16} & 1\,395.10 & {\bf 201.51} & 179.90 & {\bf 97.72} \\ \hline
			\end{tabular}
		\end{subtable}
		\begin{subtable}{1\linewidth}
			\centering
			\caption{After documents are removed}
			\label{tab:update_idx_delete}
			\begin{tabular}{|c|c|l|c|l|c|l|}
				\hline
				\multirow{2}{*}{\textbf{\begin{tabular}[c]{@{}c@{}}No.\\ Documents\end{tabular}}} & \multicolumn{2}{c|}{\textbf{Inverted Index}} & \multicolumn{2}{c|}{\textbf{Vocabulary}} & \multicolumn{2}{c|}{\textbf{PoS Index}} \\ \cline{2-7} 
				& \textbf{Update} & \textbf{Rebuild} & \textbf{Update} & \textbf{Rebuild} & \textbf{Update} & \textbf{Rebuild} \\ \hline
				\multicolumn{1}{|l|}{500} & \multicolumn{1}{r|}{{\bf 1.08}} & \multicolumn{1}{r|}{122.36} & \multicolumn{1}{r|}{1\,129.35} & \multicolumn{1}{r|}{{\bf 198.35}} & \multicolumn{1}{r|}{91.33} & \multicolumn{1}{r|}{91.91} \\ \hline
				\multicolumn{1}{|l|}{1000} & \multicolumn{1}{r|}{{\bf 1.38}} & \multicolumn{1}{r|}{123.42} & \multicolumn{1}{r|}{1\,135.08} & \multicolumn{1}{r|}{{\bf 199.20}} & \multicolumn{1}{r|}{92.00} & \multicolumn{1}{r|}{92.88} \\ \hline
				\multicolumn{1}{|l|}{2500} & \multicolumn{1}{r|}{{\bf 1.51}} & \multicolumn{1}{r|}{126.64} & \multicolumn{1}{r|}{1\,160.26} & \multicolumn{1}{r|}{{\bf 195.09}} & \multicolumn{1}{r|}{93.15} & \multicolumn{1}{r|}{94.52} \\ \hline
				\multicolumn{1}{|l|}{5000} & \multicolumn{1}{r|}{{\bf 1.60}} & \multicolumn{1}{r|}{129.76} & \multicolumn{1}{r|}{1\,202.71} & \multicolumn{1}{r|}{{\bf 203.01}} & \multicolumn{1}{r|}{96.03} & \multicolumn{1}{r|}{97.40} \\ \hline
			\end{tabular}
		\end{subtable}
	\end{table}
	
	Figure~\ref{fig:search_performance} presents the mean time for extracting the top-k documents. Tests are performed on Corpus~7 with $k=20$. After each search, the database cache and buffers are cleared so that the comparison is accurate. MongoDB is from 86\% faster than PostgreSQL for one term-search to over 50\% faster for five terms.
	
	Table~\ref{tab:create_idx} presents mean text cleaning and index construction times, as index construction is done separately in MongoDB. MapReduce functions were developed to further improve performance. Our results show that text cleaning and index creation is improved by 94\% with MongoDB (Figure~\ref{fig:text_preprocessing}). Moreover, index update is an important feature in a system where new documents are added or deleted. We use new corpora of 500 to 5,000 articles from Corpus~5 to test this feature in MongoDB. For comparison purposes, for each operation, we tested the performance of updating and rebuilding the entire index. Updating the inverted index  and the PoS index (Table~\ref{tab:update_idx_add}) works fast if the number of added documents is small, but time performance shifts for bigger corpora. Then, it is better to rebuild the entire index. If documents are deleted, it is faster to rebuild the inverted index (Table~\ref{tab:update_idx_delete}). Little improvement is seen between updating and rebuilding the PoS index (Table~\ref{tab:update_idx_delete}) when documents are deleted. Concerning vocabulary, it is faster to rebuild the entire index than to update it, because the IDF must be recomputed for each element in the collection (Tables~\ref{tab:update_idx_add} and \ref{tab:update_idx_delete}).
	
	\pgfplotsset{height=0.70\columnwidth, width=1.00\columnwidth}
	\begin{figure}[hbt]
		\centering
		\begin{subfigure}{.5\textwidth}
			\begin{tikzpicture}[]
			\begin{axis}[
			bar width=3pt,
			legend style={anchor=north west, legend cell align=left, at={(0.0,1.0)}},
			ybar = 0pt,
			xmin = 0,
			xmax = 11,
			xtick \empty,
			extra x ticks  = {1,2,3,4,5,6,7,8,9,10},
			extra x tick labels = {100k,200k,300k,400k,500k,600k,700k,800k,900k,1000k},
			x tick label style={rotate=45, anchor=east},
			xtick align=inside,
			xlabel={No. tweets},
			ytick \empty, 
			extra y ticks = {0,5000,10000,15000,20000,25000,30000,35000},
			extra y tick labels = {0,5k,10k,15k,20k,25k,30k,35k},
			ylabel={Time in seconds}
			]
			\addplot+[color=blue] table [x=tweets, y=1_thread]  {text_cleaning_threads.txt};
			\addlegendentry{1 Thread}
			\addplot+[color=red] table [x=tweets, y=12_threads]  {text_cleaning_threads.txt};
			\addlegendentry{12 Threads}
			\end{axis}
			\end{tikzpicture}
			\caption{Text cleaning comparison}
			\label{fig:text_cleaning_threads}
		\end{subfigure}%
		\begin{subfigure}{.5\textwidth}
			\begin{tikzpicture}[]
			\begin{axis}[
			bar width=3pt,
			legend style={anchor=north west, legend cell align=left, at={(0.0,1.0)}},
			ybar = 0pt,
			xmin = 0,
			xmax = 6,
			ymin = 0,
			ymax = 8,
			xtick \empty,
			extra x ticks  = {1,2,3,4,5},
			extra x tick labels = { 1, 2, 3, 4, 5 },
			xtick align=inside,
			xlabel={No. terms},
			ytick \empty, 
			extra y ticks = {0,1,2,3,4,5},
			extra y tick labels = {0,1,2,3,4,5,6,7,8},
			ylabel={Time in seconds}
			]
			\addplot+[color=blue] table [x=words, y=1_node]  {search_tweets.txt};
			\addlegendentry{1 Node}
			\addplot+[color=red] table [x=words, y=5_nodes]  {search_tweets.txt};
			\addlegendentry{5 Nodes}
			\end{axis}
			\end{tikzpicture}
			\caption{Search comparison}
			\label{fig:search_tweets}
		\end{subfigure}
		\begin{subfigure}{.5\textwidth}
			\begin{tikzpicture}[]
			\begin{axis}[
			bar width=3pt,
			legend style={anchor=north west, legend cell align=left, at={(0.0,1.0)}},
			ybar = 0pt,
			xmin = 0,
			xmax = 12,
			xtick \empty,
			extra x ticks  = {1,2,3,4,5,6,7,8,9,10,11},
			extra x tick labels = {0.1M,0.5M,1.0M,1.5M,2.0M,2.5M,3.0M,3.5M,4.0M,4.5M,5.0M},
			x tick label style={rotate=45, anchor=east},
			xtick align=inside,
			xlabel={No. tweets},
			ytick \empty, 
			extra y ticks = {0,500,1000,1500,2000,2500},
			extra y tick labels = {0,500,1\,000,1\,500,2\,000,2\,500},
			ylabel={Time in seconds}
			]
			\addplot+[color=blue] table [x=tweets, y=1_node_voc]  {voc_build.txt};
			\addlegendentry{1 Node}
			\addplot+[color=red] table [x=tweets, y=5_nodes_voc]  {voc_build.txt};
			\addlegendentry{5 Nodes}
			\end{axis}
			\end{tikzpicture}
			\caption{Vocabulary index}
			\label{fig:voc_build}
		\end{subfigure}%
		\begin{subfigure}{.5\textwidth}
			\begin{tikzpicture}[]
			\begin{axis}[
			bar width=3pt,
			legend style={anchor=north west, legend cell align=left, at={(0.0,1.0)}},
			ybar = 0pt,
			xmin = 0,
			xmax = 12,
			xtick \empty,
			extra x ticks  = {1,2,3,4,5,6,7,8,9,10,11},
			extra x tick labels = {0.1M,0.5M,1.0M,1.5M,2.0M,2.5M,3.0M,3.5M,4.0M,4.5M,5.0M},
			x tick label style={rotate=45, anchor=east},
			xtick align=inside,
			xlabel={No. tweets},
			ylabel={Time in seconds}
			]
			\addplot+[color=blue] table [x=tweets, y=1_node_ne]  {ne_build.txt};
			\addlegendentry{1 Node}
			\addplot+[color=red] table [x=tweets, y=5_nodes_ne]  {ne_build.txt};
			\addlegendentry{5 Nodes}
			\end{axis}
			\end{tikzpicture}
			\caption{NE index}
			\label{fig:ne_build}
		\end{subfigure} %
		\caption{Index construction comparison}
	\end{figure}
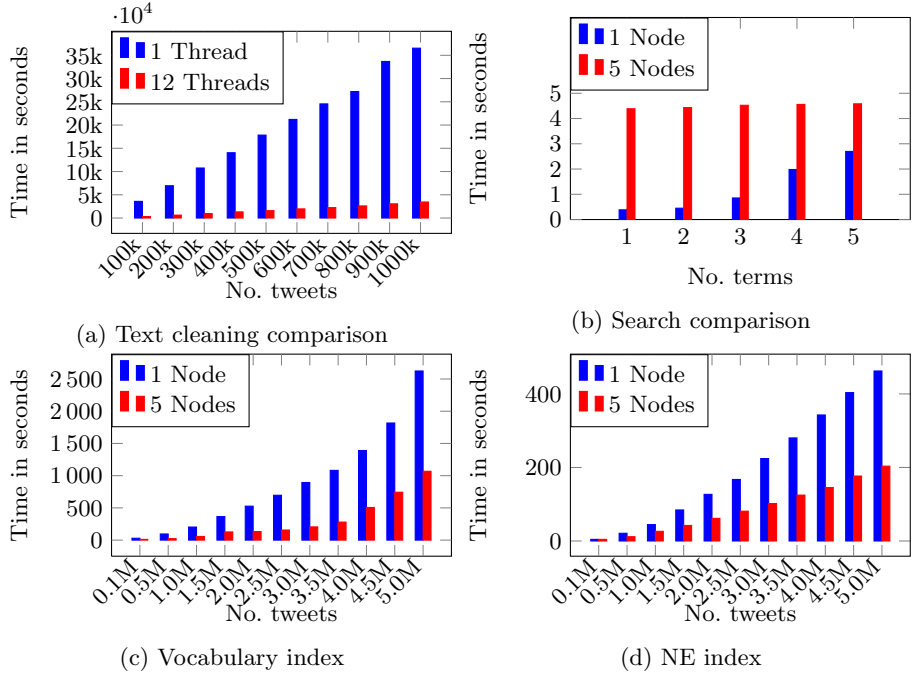 
	
	\subsection{Twitter Corpus Experiments}\label{sec:twitterxp}
	This set of experiments is carried out using one machine with the following hardware configuration: 12 GB RAM and 3 CPU with 4 2.6 GHz cores. We choose this hardware configuration to prove that our architecture does not require specialized hardware to have good time performance. We work on 5,000,000 tweets in these experiments.

	Figure~\ref{fig:text_cleaning_threads} presents the results obtained when using a multithreading architecture. The improvement obtained from switching from a single thread to a 12-thread implementation is 90\%, lowering preprocessing time by a factor of 10. We can observe that the number of nodes used by MongoDB directly impacts performance and enhances response time, especially for large numbers of tweets. The construction time of the vocabulary index improves significantly, by over 59\% (Figure~\ref{fig:voc_build}). The same happens with the named entities index, with an improvement over 40\% (Figure~\ref{fig:ne_build}). Keyword search performance remains constant when we scale the database horizontally (Figure~\ref{fig:search_tweets}).
	
	\subsection{Scientific Articles Corpus Experiments}
	
	This set of experiments uses the scientific corpus and is carried out using the same hardware configuration as in Section~\ref{sec:twitterxp}. These experiments are designed to test the  time performance for constructing the vectorization matrices and extracting topics. Figure~\ref{fig:vectorization} displays  construction time for four different vectorization matrices, namely TF, TF$_n$, TF*IDF and Okapi BM25. The best performance is obtained by the TF$_n$ vectorization matrix because all the information exists in the \emph{vocabulary} index. TF*IDF and Okapi BM25 vectorizations are slower because they must be computed for each element during matrix construction. The second set of tests presents the performance time of extracting topics from the entire corpus (\ref{fig:tm_comparison}). LSI is faster then LDA and HDP by a factor of 21 and 13,  respectively. NMF achieves the best performance.
	
	\pgfplotsset{height=0.70\columnwidth, width=1.00\columnwidth}
	\begin{figure}[hbt]
		\centering
		\begin{subfigure}{.5\textwidth}
			\centering
			\begin{tikzpicture}[]
			\begin{axis}[
			bar width=3pt,
			legend style={at={(0.5,1.0)}, anchor=south,legend columns=2},
			ybar = 2pt,
			xmin = 0,
			xmax = 5,
			xtick \empty,
			xtick align=inside,
			extra x ticks = {1,2,3,4},                    
			extra x tick labels = {5K,10K,15K,20K},
			x tick label style={rotate=45, anchor=east},
			xlabel={No. documents},                    
			ylabel={Time in seconds}
			]
			\addplot+[color=blue] table [x=test, y=count_mean ]  {vectorization_times.txt};
			\addlegendentry{$TF$}
			\addplot+[color=red] table [x=test, y=tfidf_mean ]  {vectorization_times.txt};
			\addlegendentry{$TF*IDF$}
			\addplot+[color=green] table [x=test, y=tf_mean ]  {vectorization_times.txt};
			\addlegendentry{$TF_n$}
			\addplot+[color=gray] table [x=test, y=okapi_mean ]  {vectorization_times.txt};
			\addlegendentry{$OkapiBM25$}
			\end{axis}
			\end{tikzpicture}
			\caption{Corpus vectorization comparison}
			\label{fig:vectorization}
		\end{subfigure}%
		\begin{subfigure}{.5\textwidth}
			\centering
			\begin{tikzpicture}[]
			\begin{axis}[                	
			bar width=3pt,
			legend style={at={(0.5,1.0)}, anchor=south,legend columns=2},
			ybar = 2pt,
			xmin = 0,
			xmax = 5,
			xtick \empty,
			xlabel={Algorithm},
			extra x ticks  = {1,2,3,4},
			extra x tick labels = {LSI, LDA, HDP, NMF},
			x tick label style={rotate=45, anchor=east},
			xtick align=inside,
			ylabel={Time in seconds}
			]
			\addplot+[color=blue] table [x=algorithm, y=count_mean ]  {tm_times.txt};
			\addlegendentry{$TF$}
			\addplot+[color=red] table [x=algorithm, y=tfidf_mean ]  {tm_times.txt};
			\addlegendentry{$TF*IDF$}
			\addplot+[color=green] table [x=algorithm, y=tf_mean ]  {tm_times.txt};
			\addlegendentry{$TF_n$}
			\addplot+[color=gray] table [x=algorithm, y=okapi_mean ]  {tm_times.txt};
			\addlegendentry{$OkapiBM25$}
			\end{axis}
			\end{tikzpicture}
			\caption{Topic modeling comparison}
			\label{fig:tm_comparison}
		\end{subfigure}	
		\caption{Topic modeling comparison}
	\end{figure}
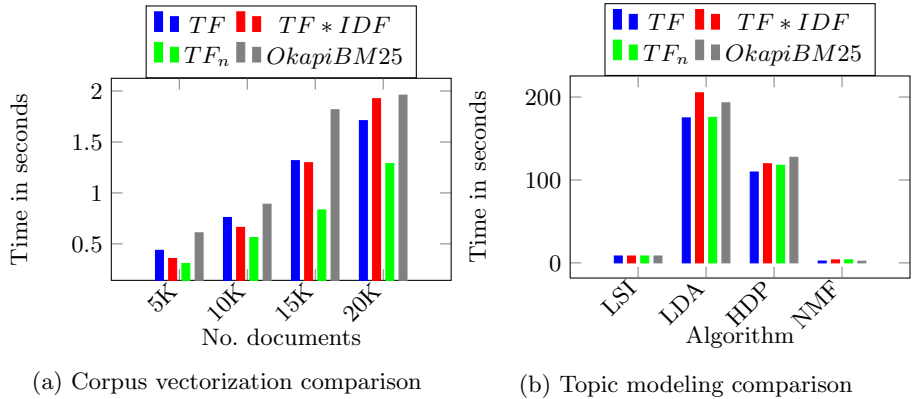
	
	\section{Conclusion}\label{sec:Conclusion}
	
	In this paper, we present a new, complete architecture for text analysis that improves search performance, minimizes storage cost through efficient document-oriented storage, and scales up horizontally and vertically. Moreover, by exploiting MapReduce to parallelize index construction and by designing new structures for indexing and decreasing the number of records stored in the database, we minimize the number of CRUD operations and further enhance performance. Finally, the algorithm we propose for extracting top-k documents for a given search phrase also considerably improves query response time.
	
	Our experimental results show that a document-oriented architecture is best-suited and improves performances when working with large volumes of text when adding documents into the database, cleaning text and constructing indexes. For all test cases, the mean time for populating the DODBMS is half that of the RDBMS. Cleaning texts and constructing inverted indexes is also faster when using a DODBMS. Although duplicates can be found inside a DODBMS, storage costs are significantly lower than with a RDBMS. A demo application that further shows the capabilities of this architecture is presented in~\cite{COTruica2016}. 
	
	In future work, we plan to add new features to our framework, such as automatic domain specific multiterm extraction, cross-language IR, word embedding and new topic models, e.g., dynamic topic modeling.
From an architectural point of view, we also want to parallelize the algorithms and use a GPU for computations. 
	
	\bibliographystyle{splncs03}
	\bibliography{adma2016}
	
\end{document}